\documentclass[letterpaper,twocolumn,10pt]{article}
\usepackage{hyperref}
\usepackage{usenix2019_v3}
\usepackage[pdf]{graphviz}
\usepackage{epsfig,endnotes}
\usepackage{color}
\usepackage[T1]{fontenc}
\usepackage{graphicx}
\usepackage[utf8]{inputenc}
\usepackage{xspace}
\usepackage{subcaption}
\usepackage{cite}
\usepackage{todonotes}
\usepackage{breakurl}

\date{}

\title{\Large \bf Co-evolving Tracing and Fault Injection with \bop}

\author {
	{ \rm Daniel Bittman} \\ UC Santa Cruz \and
	{ \rm Ethan L. Miller} \\ UC Santa Cruz \and
	{ \rm Peter Alvaro} \\ UC Santa Cruz
}

\newcommand{\paa}[1]{{\textcolor{red}{[[#1 -- paa]]}}}

\newcommand{\bop}{Box of Pain\xspace}
\newcommand{\ptrace}{\texttt{ptrace}\xspace}
\newcommand{\eex}[1]{\texttt{#1$\uparrow$}}
\newcommand{\een}[1]{\texttt{#1$\downarrow$}}

\newcounter{enumtight} 

\begin{document}

\maketitle


\begin{abstract}

Distributed systems are hard to reason about largely because of uncertainty about what may go
wrong in a particular execution, and about whether the system will mitigate those faults.  Tools that
\emph{perturb} executions can help test whether a system is robust to faults,
while tools that \emph{observe} executions can help better
understand their system-wide effects. We present \bop, a tracer and fault injector for
unmodified distributed systems that addresses both concerns by interposing at the system call level
and dynamically reconstructing the partial order of communication events based on causal
relationships. \bop's lightweight approach to tracing and focus on simulating the \emph{effects}
of partial failures on communication rather than the failures themselves sets it apart from other
tracing and fault injection systems. We present evidence of the promise of 
\bop and its approach to lightweight observation and perturbation of
distributed systems.

\end{abstract}

\section{Introduction}

Distributed systems are all around us and yet are riddled with bugs.
This should make us uneasy even if it comes as no surprise.
The space of possible executions of a distributed system is exponential
in the number of communicating processes and in the number of messages, making it difficult to build
confidence that distributed programs of even modest complexity are free from errors.
Tools that
require painstaking instrumentation and fine-grained control of runtime systems, including both
\emph{bug finding} approaches such as software model checking~\cite{verisoft, cmc} and \emph{debugging} approaches such
as deterministic replay~\cite{chen-replay,odr}, have made few inroads into distributed systems software quality
methodologies.  Instead, the field is dominated by incomplete approaches based on testing, which
can be effective at finding bugs
but cannot rule them out.

In the testing community, there is increasing interest in light-weight techniques for
\emph{observing} and \emph{perturbing} executions during integration tests, such as call
graph tracing~\cite{dapper} and targeted~\cite{fit} or random~\cite{chaos} fault injection.  
These techniques make it possible to
better cover the space of possible executions (e.g., by driving the system into rare cases triggered
by events like machine crashes and network partitions) and better understand such events'
system-wide effects.  Better still, they impose only modest overheads,
allowing observability and resiliency to be built up in a pay-as-you-go fashion.

Unfortunately, these ostensibly lightweight techniques often require instrumentation at the application
layer (e.g., propagating annotations to downstream calls or identifying
fault interposition points), a process that must be repeated for each application. Existing tracing 
and fault injection techniques tend to be coarse-grained, leading to low-fidelity signals (e.g.
call graphs whose nodes represent service endpoints) and high-overhead experiments (e.g. modeling
crash faults by rebooting servers). Moreover, since tracing and fault
injection have evolved separately, there is often an impedance mismatch between them.  For
example, a fine-grained fault injection system is of little use if the granularity of the
tracing system is too coarse to interpret the effects of the experiments.



Our philosophy on tracing and fault infrastructure is three-fold. First, a distributed system in
which independent nodes communicate via message passing will manifest \emph{any} 
fault~\footnote{We assume the \emph{omission} failure model and are concerned only
with \emph{distributed} bugs---i.e., those that could not be discovered using single-site fault injection tools.}
as the
\emph{absence} of messages (or explicit error such as timeout). To
understand the effects of these phenomena on the processes that witness them, then, a fault injector
need only focus on removing communication edges in an execution graph; thus, we can make use of a
tracing framework that focuses on reconstructing communication graphs and partial orders.
Second, we believe (and will
provide evidence) that although the space of possible executions of a distributed system is
exponentially large in the number of events, in practice some executions are significantly more
likely than others; thus, even if an understanding of a system is based on witnessing schedules of
executions, we can bound the number of schedules we are likely to see.
Third, tracing and fault-injection should \emph{co-evolve}---tracing is
necessary to inform and perform targeted fault-injection, which can only perturb events in a language that is
defined \emph{by} the tracing infrastructure itself; thus, economy of mechanism outweighs
separation of concerns.

We are building a tracing and fault injection system, \bop, which embodies our philosophy.
\bop witnesses a schedule of a distributed system execution by tracing at
the system-call level and uses those system calls to reconstruct the communication
graph of the system. We argue that this interposition point is not only effective at faithfully
capturing the communication pattern between threads (which constitutes an adequate fault surface),
but that it also manages the trade-off between generality, ease of use (as systems need not be
instrumented manually), and understanding of application-level semantics. We discuss how \bop is
able to effectively trace and inject faults in a distributed system because, while the space
of possible executions is large, we often need only a small representation of the whole system to
find bugs~\cite{alloybook,simpletesting}, but also because these possible different execution
schedules will often be consistent with the same partial order, and so are effectively the ``same''
execution, moving a theoretically intractable problem into the practical realm.

\section{Background}


Unlike traditional model checkers that identify bugs in \emph{specifications}, software model
checkers (SMCs)~\cite{verisoft,cmc} systematically explore the state space of actual
\emph{implementations} via fine-grained control of a program's execution schedule, and backtracking
as necessary.  When a bug is identified in such a concurrent system, it is often challenging to
reproduce when debugging. Deterministic replay systems~\cite{chen-replay,odr} make this possible
by recording \emph{traces} that capture non-deterministic inputs or events and then, much like SMCs,
controlling the runtime schedule during replay to ensure that the same events occur in the same
order.  Like these ``heavyweight'' techniques, we want to work with
arbitrary, unmodified systems by instrumenting relatively low in the stack.  However, fine-grained
scheduling is costly to run and implement, and is overkill for the tasks of distributed
tracing and fault injection.

Lightweight approaches to observing distributed executions based on call graph
tracing~\cite{jaeger,ldfi-socc,opentracing} have gained a great deal of popularity in recent years,
and a number of businesses are devoted to the collection and analysis of call graph
traces~\cite{honeycomb,lightstep}.  These observability infrastructures, based on Google's
Dapper~\cite{dapper} require modifications to application code in order to propagate trace
annotations (unique identifiers and other adornments) that are attached to incoming service requests
to downstream service calls.  This boilerplate, while relatively straightforward to write, imposes a
significant burden on the application programmer and must be repeated for each application.  While we
wish to provide value without requiring work on the part of the application programmer, we would
nevertheless like to be able to reconstruct this \emph{application-level} signal from
instrumentation lower in the stack.

The distributed resiliency community has long advocated combining testing methodologies with fault
injection~\cite{chaos,orchestra,fate} to increase confidence that ostensibly fault-tolerant
programs operate correctly under the (rare in practice) fault events that they were designed to
mitigate.  Although as discussed fault injection infrastructures are often used in concert with
tracing, they have tended to develop as separate concerns.  A stated goal of \bop is
to coevolve these concerns.

The data management community has used data lineage~\cite{tiresias, provenance-db,
	fo-games, constraint-games, whynot-sdn} to explain query answers in much the same way that the
resilience community uses call graph tracing to explain distributed executions.  Lineage-driven
fault injection~\cite{molly,ldfi-socc}, a bug-finding technique that we will discuss further in
Section~\ref{sec:future}, directly uses explanations of system outcomes (formal data lineage
or execution traces) to 
automate fault injection experiments.
\bop was designed to integrate
tightly with such a bug finder, providing it with traces as performing the fault injection
experiments that it suggests.

\section{A Partial Argument of a Partial Order}

When tracing a distributed system, we often have two options: build tracing infrastructure into the
application during initial development, which requires difficult forethought, or build it in afterwards,
which requires a significant engineering effort that is often avoided unless necessary. Instead, if
we could trace a system at a level that provides sufficient signal to reconstruct communication
we could circumvent the complexity of kernel-level tracing and the overhead of application-level instrumentation.
We propose tracing at the system-call level, as this
is transparent to the application, can run on an unmodified system, allows easy experimentation by
simulating a system on one machine, and can still derive sufficient signal to be useful for fault
injection and collection of rich system traces, as we see in this section.

One significant consequence of tracing system calls is that the tracer will see a
schedule of events with little ordering among them. While each observed event
on a per-thread basis is ordered with respect to other events in that thread, there are no immediate constraints on event ordering
\emph{between} threads.
Although the tracer sees a sequentially consistent execution
consistent with the true partial order of events, it cannot determine a richer partial order
beyond this independent collection of total orders from witnessing schedules alone.

To understand the communication structure of a program as well as to inject faults,
however, more than just this weak schedule is needed.  Fortunately, since we know the
\emph{semantics} of the system calls, we can use their meaning to glean more information from them
than we could if we strictly observed them in a particular schedule. For example, a given (successful) call to
\texttt{accept} cannot return until a paired call to \texttt{connect} is made, or a (successful) call to
\texttt{read} on a
socket cannot return until a causally-paired call to \texttt{write} is made.

The additional ordering available to us from observing socket system calls and tracking connection
and data transfer is exactly the communication pattern between the threads in the system.  We can
use that communication pattern to derive \emph{happens-before}, which characterizes the constraints
between events of different threads, thus enabling fine-grained, targeted fault injection that can specify ``when'' in a
distributed execution to inject faults based on the communication pattern rather than wall-clock time.
Furthermore, this pattern can be derived \emph{during} execution (as opposed to reconstructed after
completion), a requirement of a fault-injection infrastructure that injects targeted faults based on
moments within a trace. The mechanism for this is described in Section~\ref{sec:tracking}.


Of course, a distributed execution might be different each time it runs, as there is inherent
non-determinism in message delivery and thread scheduling. While the theoretical  behavior of a system is
characterized by a collection of all possible partial orders of events, for the purposes of fault
injection it might be sufficient to collect only a limited number of such schedules,
\emph{especially} if some schedules are more likely than others. The intuition behind schedules
having different probabilities is straight-forward: the most significant source of non-determinism
in a distributed system is the real-time order of events between threads. However, if we recognize
that we will observe the events in \emph{some} sequentially consistent order, and we know that the
events per-thread are totally ordered, then for the purposes of comparing two schedules we can
ignore the actual order we observe the events in \emph{as long as} both schedules are consistent
with the same partial order.

Thus, we are left with collecting schedules of distributed systems that are \emph{truly different}
in their communication patterns and behavior. This dramatically reduces the space of executions,
down from exponential in the number of events and number of threads to the number of 
valid communication patterns given a single input (which, depending on the system, might still be
large).  We hypothesize (and provide initial evidence) that most of the time, given a consistent
input, a distributed system will often produce similar partial orders, thus allowing us to construct
a representation of the system's behavior with a small set of runs and use that to inject faults.

If this is true, we open up a wealth of possibilities, because we can then trace a distributed
system and inject faults repeatedly, witnessing ``good'' and ``bad'' executions, and adjusting our
fault injection over time on a \emph{real} system with \emph{no} manual instrumentation. This is the
goal of \bop---to use the application-level signal we derive in a generic, low-overhead way to
inform the decisions of bug-finding frameworks and thus fully automate tracing, fault-injection, and
bug-finding.

\section{Box of Pain}

\bop has three components: a \emph{tracer}, a \emph{tracker}, and an \emph{injector}. These
components all operate together, watching a distributed execution unfold. When run in a loop,
\bop will determine if the execution has been seen before, allowing it to build a collection
of traces that together characterize the relevant behaviors of the system. Optionally, \bop can be run
with a failure specification that indicates precisely which events to interrupt or modify as part of
fault injection (which we discuss in Section~\ref{sec:bopfi}).


\subsection{Tracing}

\bop operates primarily through the use of \ptrace, a system call that allows a process to perform
introspection on another process~\cite{ptrace}. Whenever a traced thread (tracee) issues a system
call, the tracee is stopped and \bop wakes up. This occurs both for system-call entry and exit, each
referred to as an event, and \bop handles each event in full before signaling the thread to resume.
Each event that \bop handles is appended to a per-thread ``event log'', and is thus in the order
that they occur for that thread. An entry-to-syscall event is indicated like \een{read}, and a
return-from-syscall event is indicated like \eex{read}.

The ultimate goal of tracing is to construct a partial order of events out of the schedule that
\bop observes. Given just a per-thread event log, we have a partial order (a collection of total
orders, one for each thread), but this partial order contains no constraints on events
\emph{among} threads. Since the communication pattern between two threads and the contraints on
ordering are equivalent in our model, we can leverage the information available in a TCP connection
to provide additional edges in the partial order for a given run.

When a socket is created, it is tracked in a per-process lookup table (in a way that keeps track of
changing file descriptors).  During a \een{bind} event, \bop reads the process's memory to determine
the address and port that the socket is being bound to.  After the subsequent \eex{accept} event, a
new socket is tracked (consistent with the semantics of \texttt{accept}). Since it is also tracing
the \texttt{connect}-ing thread, it will see the resultant \een{connect} and \eex{connect} events,
the first of which provides enough information for \bop to decide which socket it is connecting to,
but not necessarily which socket returned by \texttt{accept} the \texttt{connect}-ing thread is
actually associated with.

\begin{figure}
\centering
\includegraphics[width=\columnwidth]{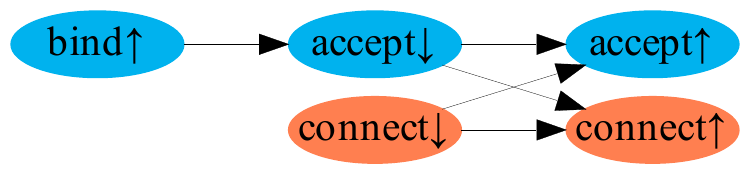}
\vspace*{-2mm}
\caption{The \emph{happens-before} relationship of \texttt{accept} and \texttt{connect} system
	calls that \bop derives. The colors indicate different threads. The \eex{connect} cannot occur before
	\een{accept} occurs, because it would result in an error otherwise.}
\label{fig:po_acc_conn}
\vspace{-2mm}
\end{figure}

To get this information, \bop issues system-calls on behalf of the tracees while handling the
\eex{connect} and \eex{accept} events.  It does this by overwriting the registers of the tracee to
point to a location known to contain a \texttt{syscall} instruction (determined during the first
event handled per-process), and setting the registers as required for the requested system call.  In
this case, the system calls are \texttt{getsockname} and \texttt{getpeername}, which provide
sufficient information to determine the end-points of the TCP stream. The resulting partial order is
shown in Figure~\ref{fig:po_acc_conn}.

For data transfer, we can use the tracked sockets to watch as TCP traffic is communicated between
end-points. When handling a \eex{write}, \bop tracks the sequence number of the stream and records
to which system call a particular range of data belongs.  When handling a \eex{read}, \bop looks
through the recorded \texttt{write} system calls to decide which writes contributed to the data
returned by the \texttt{read}, thus deriving an order based on the commuication pattern of data
transfer.  Note that one \texttt{read} can get data from multiple \texttt{write}s and one
\texttt{write} can contribute to multiple \texttt{read}s, or it can be a one-to-one relationship.


While many of these system calls have variants (\texttt{send} instead of \texttt{write}, or
\texttt{accept4} instead of \texttt{accept}), the variants are similar enough
that they need little additional processing. One exception is the calls \texttt{sendto} and
\texttt{recvfrom}, however these calls are infrequently used for TCP communication.

Finally, while \bop traces a distributed system as a set of threads in processes on a single node,
we see it as merely an engineering effort to extend the tracing infrastructure to multiple
nodes. A single tracer process can run on each node, forwarding event information to a single,
unified tracker node that processes schedules and computes partial orders.

\subsection{Tracking}
\label{sec:tracking}

The tracing infrastructure builds a trace of a distributed system that consists of a
per-thread event log, where each event can have multiple parents (as derived by the communication
pattern). The trace can be serialized and viewed as a PDF, showing the communication
pattern. However, executions may differ between runs, and if we want to be able to get an idea of
the ``true'' communication pattern between nodes in a system, we'll need to observe many of the
possible schedules.

\bop faciliates this by allowing previously collected traces to be reloaded into
memory before tracing a new run.  During execution, \bop tries to track each loaded run by
comparing the event that just occurred in the new trace to the ``next'' event in each loaded run.
``Next'' here means, ``for this thread, what was the next witnessed event''. For example, if
thread $T$ records events $e$ followed by $e'$, then a run is said to be ``followed'' if thread $T$
is witnessed executing those events in the same order, even if another thread executes some other
event in between $e$ and $e'$. When a particular run cannot be followed, \bop
stops tracking it. If all loaded runs are not followed, \bop finishes tracing the
execution and serializes the trace as before. If instead, at the end of the execution, a run
is followed, \bop does not serializing the current trace since it is equivalent to the
followed run.

\subsection{Fault Injection}
\label{sec:bopfi}

When running \bop on a distributed system, we can provide a fault specification that describes which
events to perturb via fault injection. Since we are considering the space of faults to include only delay
(possibly infinite) and explicit errors, \bop allows the simulation of both.
The tracing infrastructure that \bop provides, and the corresponding derivation of a
partial order that reflects the communication structure of the system, faults in \bop can be thought
of in terms of specifying a particular event, which allows bug-finding software to consider faults
in terms of ``after thread $T$ does $x$ and thread $T'$ does $y$'', which improves how
\emph{targeted} faults can be.

Delay is simulated by simply ``pausing'' a process by stopping and resuming it after an
amount of time (or delaying it indefinitely). This kind of fault is the simplest to inject,
as all that is required for the fault specification is a particular event identification.
If a followed run contains such a specification, \bop will
pause the process when it observes it trying to execute that event.

Explicit error manifests more directly as \bop changing the return value (or parameters)
during a system call. For example, interrupting a connection between process $A$ and $B$ is done by
changing the \texttt{socket} argument to \texttt{connect} to $-1$, thus ensuring that the connection
fails, while changing the return value to a specified error code (such as \texttt{-ECONNREFUSED}).

We can map ``real'' events into our simulation space via a combination of manifesting explicit
errors, pausing processes, or silently dropping communications. A lengthy garbage-collection
pause can be emulated by pausing a process for some time, while a machine crash can be
emulated by either stopping a process indefinitely, restarting the process, or dropping all messages
after a point in the partial order. Network partitions are similar; we can observe the
destinations of messages and drop them (either silently or via an error) if we simulate
them crossing a network partition. Later, healing the network partition can manifest as removing
those fault rules.

\section{Preliminary Experiments}

To provide an initial look at how many unique runs are generated by some small distributed systems,
we ran the Redis key/value store~\cite{redis} under \bop, and counted how many unique runs were generated varying
different parameters: number of commands performed, and number of clients. We then re-ran our
two-client scenario, but simulated network congestion by randomly forcing writes to act as if there were full
TCP buffers, randomly causing writes to only \emph{actually} write half or less than the system
call originally would have. This was done by changing the \texttt{count} parameter of the
\texttt{write} system call, reducing it to a lower value than its original value.

\begin{figure}[t]
\centering
\includegraphics[width=\columnwidth]{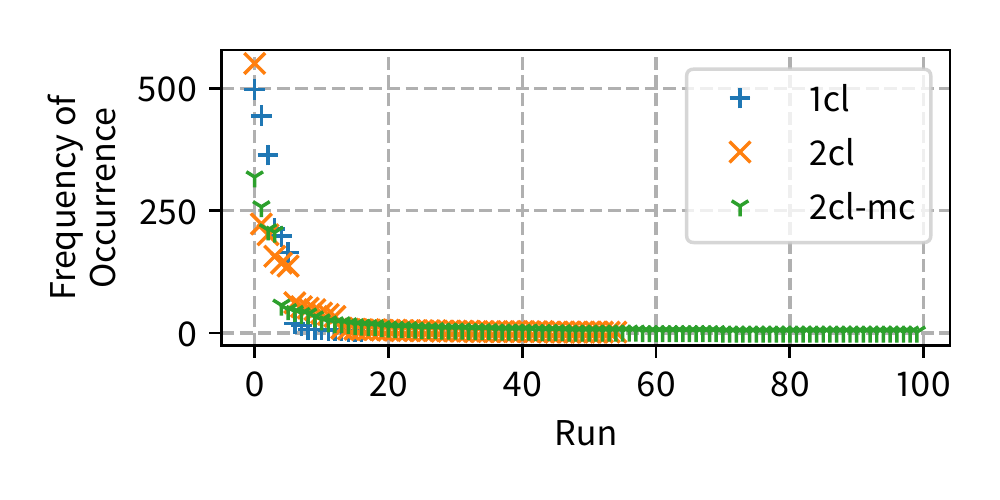}
\vspace*{-10mm}
\caption{Distribution of runs for one client (1cl), two clients (2cl), and many commands (2cl-mc).}
\label{fig:dist}
\vspace*{-5mm}
\end{figure}

\begin{figure}[t]
\centering
\includegraphics[width=\columnwidth]{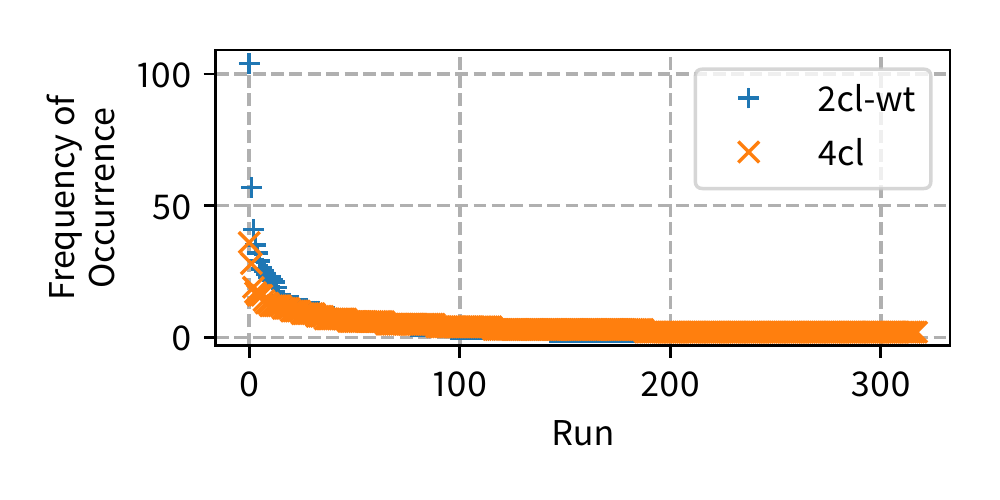}
\vspace*{-10mm}
\caption{Distribution of runs for two clients with full-TCP-buffer simulation (2cl-wt) and four
	clients (4cl). While the distribution is wider than Figure~\ref{fig:dist}, it follows a similar
shape.}
\label{fig:dist2}
\end{figure}

Figures~\ref{fig:dist} and~\ref{fig:dist2} show the distribution of runs for executions by varying
the parameters as described. We ran \bop for two thousand iterations on each configuration, and the
graphs show the runs that make up 99\% of the resulting schedules.  Each client executed a simple
\textsf{GET} request for all experiments except the experiment where we increased the number of
commands executed by each client (2cl-mc), in which case each client executed four \textsf{GET} and
\textsf{SET} commands. In all cases, the distribution rapidly drops after a high initial value,
indicating that the majority of runs manifests as one of a few schedules, and, while there is a long
tail, we can understand much of the system behavior without needing an intractable number of
schedules. Interestingly, increasing the number of commands issued by each client did not
dramatically impact the width of the distribution, indicating that bugs arising from complex series
of interactions would be easier to find. While the results in Figure~\ref{fig:dist2} have longer
tails, both of these have the same shape and 99\% of runs fall within approximately 300 already
known schedules after 2000 iterations (which took well under an hour to generate).


\section{The Future for \bop}
\label{sec:future}

As excited as we are to introduce \bop and to argue its potential, we must admit that we have barely begun using it.
In this section we describe what comes next, from first steps to a (we believe) far-reaching vision.

\bop combines two concerns---tracing and fault injection---that are typically considered separate.
Before tackling our larger ambitions, we plan to demonstrate its efficacy for both independent
tasks.  It remains to be shown that it is possible to extrapolate from our low-level traces
something akin to the application-level signal provided by call graph tracing. Tracing a large-scale
microservice-based application with \bop and showing that the call graphs (e.g., obtained using
Zipkin) could be inferred from our traces would provide evidence that technologies like \bop could
obviate the need for painstaking application-level instrumentation in some cases.  Similarly, we
will compare \bop with the state-of-the-art in distributed fault injection.  While most of these
approaches focus on triggering~\cite{chaos} or simulating~\cite{fit} fault events such as machine
crashes, I/O errors, memory pressure and corruption, system load, and so on, our approach focuses
instead on simulating the \emph{observable effects} of such faults from the perspective of other
processes with which they communicate.  We expect that this much smaller fault surface will be
sufficient to uncover bugs in fault tolerance logic and much more efficient at doing so.

From the beginning our intention has been to use \bop in a tight loop with a
trace-driven bug finder such as LDFI.  To date, LDFI has shown promise in verifying
\emph{protocols}~\cite{molly} as well in finding bugs in large-scale, microservice-based
applications~\cite{ldfi-socc}.  In the former, programs must be specified in a custom
relational logic language (similar to solvers such as Alloy~\cite{alloybook}), limiting
applicability to real-world systems. In the latter, the systems must already be instrumented to
support call graph tracing and fine-grained fault injection.  By addressing both concerns at the
system level, \bop promises to open up the LDFI approach to arbitrary, uninstrumented systems,
including distributed data managment systems, configuration services, and messages queues.


\section{Conclusion}

In our field there are a great many things that are possible in theory but impractical in
practice---so much so that the idea is a cliche. However, it is a rare day on which we learn that
something which is not possible in theory is not merely possible, but \emph{practical} in practice.
\bop's design for tracing and tracking is predicated on the idea that fault injection naturally fits
with tracing; after all, if you \emph{want} targeted fault injection, what better place to do it than
in the tracer itself? The coevolution of these technologies will open a wealth of possibilities that
we can make use of to further close the gap between the bugs we can easily find and the bugs we
could find if only we had sufficient tracing, a bug-finder, and infrastructure support---all without
the need for tracing forethought or huge engineering efforts. We have initial evidence that not only
is it possible to trace a distributed system at the system call level and recover
\emph{happens-before} such that we can decide and target faults to inject, but we can do this
without the non-determinism becoming intractible. We are excited to keep exploring this work, and
evaluating more complex systems, looking for bugs, and further evaluating our hypothesis.

\bibliographystyle{plain}
\bibliography{main.bib,bib/csrg.bib}

\end{document}